\documentclass[aps,pre,reprint,superscriptaddress]{revtex4-2}
\usepackage{bm}
\usepackage{amssymb}
\usepackage{pifont}
\usepackage{amsmath}
\usepackage{graphics}
\usepackage{graphicx}
\usepackage{epsfig}
\usepackage{pstricks}
\usepackage{pst-plot}
\usepackage{pst-slpe}
\usepackage{latexsym}
\usepackage{bm}
\usepackage{url}
\usepackage{pifont}
\usepackage{amsmath}
\usepackage{amsfonts}
\usepackage[margin=2.5cm]{geometry}
\usepackage{subfigure}
\usepackage{placeins}
 \newcommand{\be}{\begin{equation}}
 \newcommand{\ee}{\end{equation}}
 \newcommand{\bea}{\begin{eqnarray}}
 \newcommand{\eea}{\end{eqnarray}}
\usepackage{lineno,hyperref}

\begin{document}

\title{Site and bond percolation in linearly distorted triangular and square lattices}
\author{Bishnu Bhowmik}
\affiliation{Department of Physics, University of Gour Banga, Malda-732103, India}

\author{Sayantan Mitra}
\affiliation{Department of Physical Sciences, Indian Institute of Science Education and Research Kolkata, Mohanpur, 741246 India}

\author{Robert M. Ziff}
\affiliation{Center for the Study of Complex Systems and Department of Chemical Engineering, University of Michigan, Ann Arbor, Michigan 48109-2800, USA}

\author{Ankur Sensharma}
\email{itsankur@ugb.ac.in}
\affiliation{Department of Physics, University of Gour Banga, Malda-732103, India}

\begin{abstract}
We investigate site and bond percolation in triangular and square lattices subjected to linear distortion. In contrast to previously studied distortion schemes that preserve lattice geometry, linear distortion dislocates regular lattice sites along a fixed direction. Nearest-neighbors of a regular lattice need to satisfy a distance-based connection criterion to remain neighbors in the linearly distorted lattice. Using extensive Monte Carlo simulations and finite-size scaling analyses, we examine how site and bond percolation thresholds vary with the distortion parameter and the connection threshold. For triangular lattices, we observe pronounced directional dependence of both site and bond percolation thresholds, as well as of the critical connection threshold. This arises from the distortion-induced anisotropic modification of nearest-neighbor separations. In particular, bond percolation exhibits nontrivial behavior that cannot be explained solely in terms of changes in the average coordination number. In contrast, square lattices remain effectively isotropic under linear distortion, resulting in identical percolation thresholds for distortions applied along different directions. Percolation thresholds in the thermodynamic limit, evaluated for a selected set of values of distortion parameter and connection threshold, confirm that the results for large finite lattices provide reliable estimates of the infinite-system behavior.
\end{abstract}
\maketitle

\section{Introduction}
Owing to its simplicity, percolation has often been described as an accessible route into modern research, requiring only minimal background in physics or mathematics \cite{Stauffer}. It has attracted considerable interest among theorists and mathematicians due to its conceptual clarity and theoretical appeal. As a result of its flexibility, percolation has found applications across a wide range of disciplines \cite{Saberi1}, including transport in porous media \cite{Sahimi,Hunt}, complex networks \cite{Callaway,Artime}, spreading of epidemic diseases \cite{Grassberger,Moore, Miller, Ziff2, Dipa2023} and forest fires \cite{Albano1, Guisoni,Sayantan1}, and problems in the social sciences \cite{StaufferSocial}. From a physicist’s perspective, percolation is particularly intriguing because it exhibits nontrivial critical behavior \cite{Stauffer,Grimmett,ChristensenMoloney,Ball, Dotsenko}.

Percolation concerns the formation of long-range connectivity across a system through the emergence of a giant cluster that spans its boundaries. In the two basic percolation models, namely site percolation \cite{Hammersley} and bond percolation \cite{Broadbent}, clusters are formed from neighboring occupied sites or occupied bonds, respectively. A site or bond is typically occupied at random with probability $p$, known as the occupation probability. As $p$ increases, clusters grow in size. At the critical occupation probability $p_{\mathrm{c}}$ (usually denoted by $p_{\mathrm{b}}$ for bond percolation), the largest cluster first connects opposite boundaries of the system. This cluster is referred to as the spanning cluster, and its emergence signals a continuous geometric phase transition. In the site-bond percolation model \cite{Sykes1964,Tarasevich}, both occupied sites and occupied bonds participate in cluster formation process. Exact percolation thresholds for some two-dimensional lattices can be derived using duality \cite{Sykes1964,Stauffer}. For example, for the triangular lattice, the site and bond thresholds are $p^{(\triangle)}_\mathrm{c} = 1/2$ and $p^{(\triangle)}_\mathrm{b} = 2\sin(\pi/18)$, respectively, while for the square lattice, the bond threshold is $p^{(\square)}_\mathrm{b} = 1/2$ \cite{Kesten}. Star--triangle transformations can be used to obtain exact thresholds for more complex planar lattices \cite{ZiffScullard,ScullardZiff}. In addition to analytical results, percolation thresholds have been extensively estimated numerically using Monte Carlo simulations and other computational methods for various lattices with usual neighborhoods \cite{SudingZiff1999,Deng,Wang,Jacobsen_2015,Manna,Ballesteros,Gonzalez,Lorenz} as well as extended or complex neighborhoods \cite{Malarz,Malarz2021,Malarz2022,Malarz2024,Xun1,Xun2,Xun3}.

Beyond the basic site and bond models, several extensions of the classical percolation framework have been explored to account for additional constraints or correlations, including explosive percolation \cite{Achlioptas}, directed percolation \cite{Takeuchi}, continuum percolation \cite{Hall1985,Mertens}, bootstrap percolation \cite{Adler},  and correlated percolation \cite{Coniglio}.
\begin{figure*}[!htbp]
	  \subfigure[]{\raisebox{-0.2cm}{\includegraphics[scale=0.9]{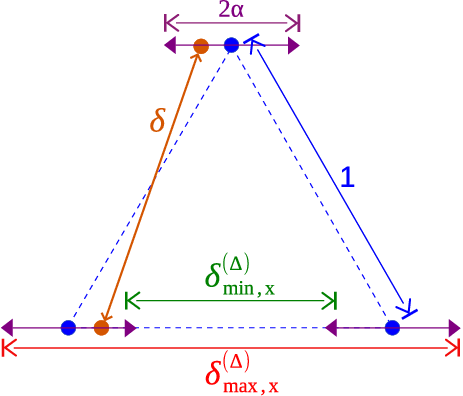}}}\hfill
      \subfigure[]{\includegraphics[scale=0.9]{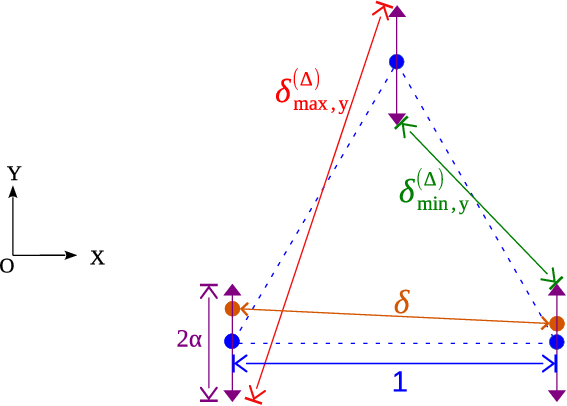}}

\caption{Mechanism of linear distortion in a triangular lattice: (a) distortion along the $x$ direction and (b) distortion along the $y$ direction. Regular lattice sites, represented by filled blue circles, are displaced within a range $2\alpha$, indicated by purple horizontal arrows in panel (a) and vertical arrows in panel (b). Filled orange circles are examples of the lattice sites after distortion. The modified nearest-neighbor distances are represented by $\delta$. The minimum and the maximum possible values of $\delta$ have been indicated by green and red arrowed lines, respectively.}
\label{fig:dist_mech_tri}
\end{figure*}

In a series of earlier works, we explored the effects of geometric distortion on percolation properties. We first studied site percolation on a distorted square lattice, demonstrating that lattice distortion significantly raises the site percolation threshold while preserving the universality class \cite{Sayantan1}. This approach was subsequently extended to three dimensions for site percolation on distorted simple cubic lattices \cite{Sayantan2}. It was also established that, despite certain similarities, percolation in distorted lattices and site–bond percolation constitute distinct models. In a later study, the role of distortion was further examined in systems with a flexible number of neighbors, where the possibility of links to extended neighbors was explored to investigate its influence on percolation behavior in both distorted square and distorted simple cubic lattices \cite{Sayantan3}. More recently, bond percolation on distorted square and triangular lattices was investigated, highlighting the combined effects of lattice geometry and distortion on bond connectivity \cite{Bishnu1}. 

In all these works, the distortion scheme followed the underlying lattice geometry: a site in a regular square, simple cubic, or triangular lattice was displaced to a random position within a square, cubic, or triangular distortion zone, respectively. In the present work, we deviate from this approach and instead employ linear distortion along either the $x$ or $y$ direction in triangular and square lattices. The cluster identification and percolation processes are simulated using the Newman–Ziff algorithm \cite{Newman1,Newman2}, as in our earlier studies.
\begin{figure*}[!htbp]
	  \subfigure[]{\includegraphics[scale=0.8]{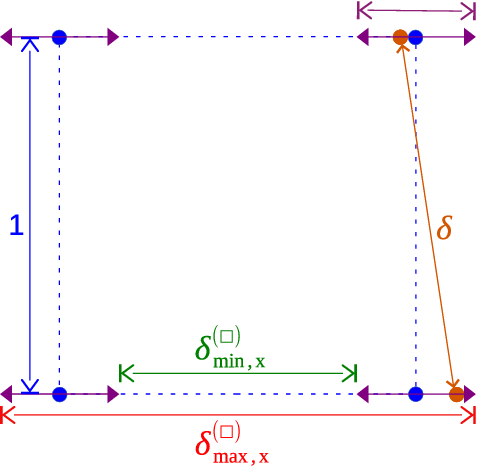}}\hfill
      \subfigure[]{\includegraphics[scale=0.8]{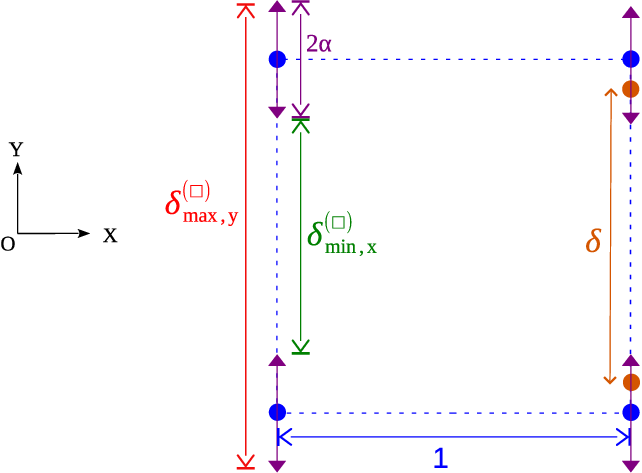}}

\caption{Mechanism of linear distortion in a square lattice: (a) distortion along the $x$ direction and (b) distortion along the $y$ direction. Regular lattice sites, represented by filled blue circles, are displaced within a range $2\alpha$, indicated by purple horizontal arrows in panel (a) and vertical arrows in panel (b). Filled orange circles are examples of the lattice sites after distortion. The modified nearest-neighbor distances are represented by $\delta$. The minimum and the maximum possible values of $\delta$ have been indicated by green and red arrowed lines, respectively.}
\label{fig:dist_mech_squ}
\end{figure*}

The remainder of the paper is organized as follows. Section~\ref{sec:model} illustrates the linear distortion scheme. In Sec.~\ref{sec:cluster}, the connection criterion is explained and the numerical protocol for cluster formation is described. Section~\ref{sec:main} presents the main results of this paper, namely the variation of site and bond percolation thresholds with distortion for triangular (Sec.~\ref{sec:tri}) and square (Sec.~\ref{sec:squ}) lattices. For the triangular lattice, results corresponding to linear distortion along the $x$ and $y$ directions are presented separately. Precise percolation thresholds in the thermodynamic limit for selected parameter values are reported in Sec.~\ref{sec:inf}. The variation of the critical connection threshold is discussed in Sec.~\ref{sec:dc}, followed by concluding remarks in Sec.~\ref{sec:summary}.
\section{The distortion mechanism}\label{sec:model}

Figs. \ref{fig:dist_mech_tri} and \ref{fig:dist_mech_squ} illustrate the mechanism of how the lattice points are distorted in a triangular and square lattice respectively. Distortion is controlled by the distortion parameter $\alpha$. Each lattice point of a regular triangular (square) lattice of lattice constant $1$ is dislocated by a random amount within a line segment of length $2\alpha$ centered at the regular lattice point. We consider two cases, as shown in Figs. \ref{fig:dist_mech_tri} and \ref{fig:dist_mech_squ}: (i) horizontal line segments, corresponding to  dislocations along the $x$ direction,  (ii) vertical line segments, corresponding to dislocations along the $y$ direction. 

Consider a site at position $(i,j)$ in the regular lattice. Under linear distortion, the site is displaced to $(i + r_x, j + r_y)$, where $r_x, r_y \in (-\alpha, \alpha)$ are independent random variables. For $x$-distortion, $r_y = 0$, whereas for $y$-distortion, $r_x = 0$.

With the lattice constant set to unity, the nearest-neighbor distance in the undistorted lattice is $1$. The distance between two displaced neighboring sites is denoted by $\delta$, which depends on the lattice geometry and the direction of distortion.

Representative examples of dislocated lattice sites are shown in Figs. \ref{fig:dist_mech_tri} and \ref{fig:dist_mech_squ} for all four cases. Note that, under distortion along the $x$ direction, the horizontal bonds retain their orientation for both lattices, while the oblique bonds of the triangular lattice and the vertical bonds of the square lattice change their orientation. Under $y$-distortion, the horizontal bonds become oblique, whereas the vertical bonds of the square lattice retain their orientation; in contrast, the oblique bonds of the triangular lattice also change their orientation.

The minimum and maximum possible values of $\delta$ of the triangular lattice for distortion along $x$ and $y$ directions are denoted by $\delta_{\mathrm{min,x}}^{\mathrm{(\triangle)}}, \delta_{\mathrm{max,x}}^{\mathrm{(\triangle)}}, \delta_{\mathrm{min,y}}^{\mathrm{(\triangle)}}, \delta_{\mathrm{max,y}}^{\mathrm{(\triangle)}}$. The corresponding quantities of the square lattice are denoted by $\delta_{\mathrm{min,x}}^{\mathrm{(\square)}}, \delta_{\mathrm{max,x}}^{\mathrm{(\square)}}, \delta_{\mathrm{min,y}}^{\mathrm{(\square)}}, \delta_{\mathrm{max,y}}^{\mathrm{(\square)}}$. These quantities are given by
\begin{equation}
\label{eq:maxmin}
\begin{aligned}
&\delta_{\mathrm{min,x}}^{\mathrm{(\triangle)}}=\delta_{\mathrm{min,x}}^{\mathrm{(\square)}}=\delta_{\mathrm{min,y}}^{\mathrm{(\square)}}=1-2\alpha,\\
&\delta_{\mathrm{max,x}}^{\mathrm{(\triangle)}}=\delta_{\mathrm{max,x}}^{\mathrm{(\square)}}=\delta_{\mathrm{max,y}}^{\mathrm{(\square)}}=1+2\alpha,\\    
&\delta_{\mathrm{min,y}}^{\mathrm{(\triangle)}}=\sqrt{1-2\sqrt{3}\alpha +4\alpha ^2},\\
&\delta_{\mathrm{max,y}}^{\mathrm{(\triangle)}}=\sqrt{1+2\sqrt{3}\alpha +4\alpha ^2}.\\
\end{aligned}
\end{equation} 

It can be seen from Figs.~\ref{fig:dist_mech_tri} and \ref{fig:dist_mech_squ} that, under $x$-distortion, the longest and shortest bonds are always horizontal. In contrast, under $y$-distortion, the maximum and minimum bond lengths occur for non-horizontal bonds. Eq.~\ref{eq:maxmin} reveals that the minimum and the maximum distances in two directions are the same for a square lattice. In contrast, they are different for a triangular lattice since $\delta_{\mathrm{min,x}}^{\mathrm{(\triangle)}} \neq \delta_{\mathrm{min,y}}^{\mathrm{(\triangle)}}$ and $\delta_{\mathrm{max,x}}^{\mathrm{(\triangle)}} \neq \delta_{\mathrm{max,y}}^{\mathrm{(\triangle)}}$. This directional dependence impacts the local connectivity and breaks the isotropy in the triangular lattice leading to distinct percolation thresholds for distortion applied along two different directions.

\begin{figure*}[!htbp]
\centering
	 {
	  \subfigure[]{\includegraphics[width=0.33\textwidth]{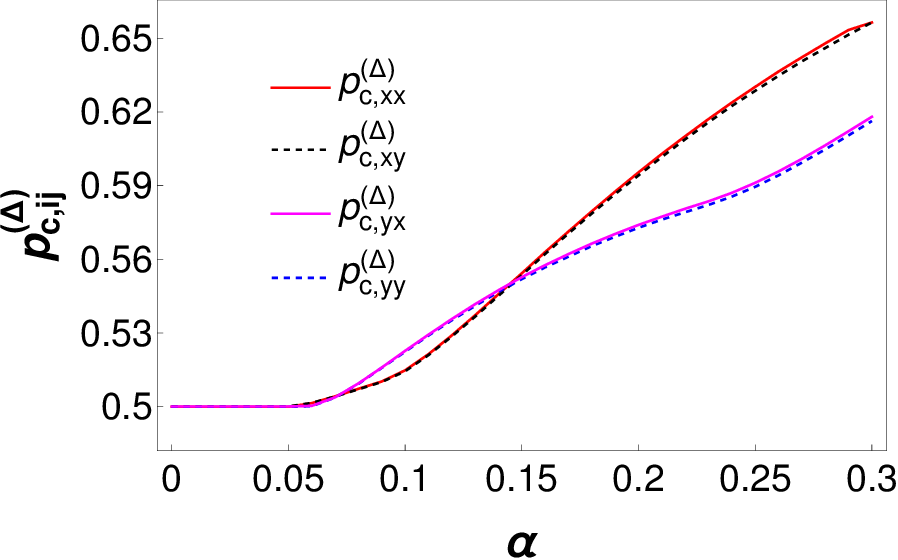}}
	  \subfigure[]{\includegraphics[width=0.34\textwidth]{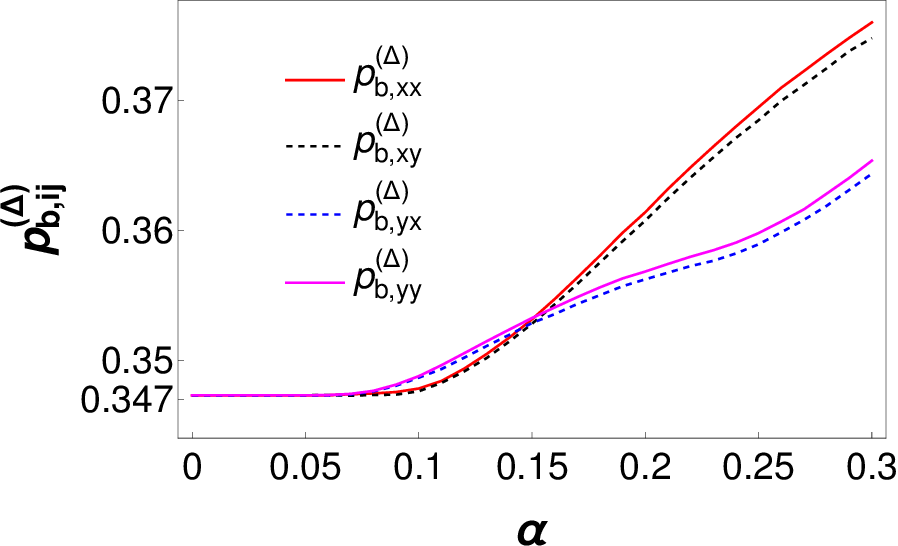}}
      \subfigure[]{\includegraphics[width=0.31\textwidth]{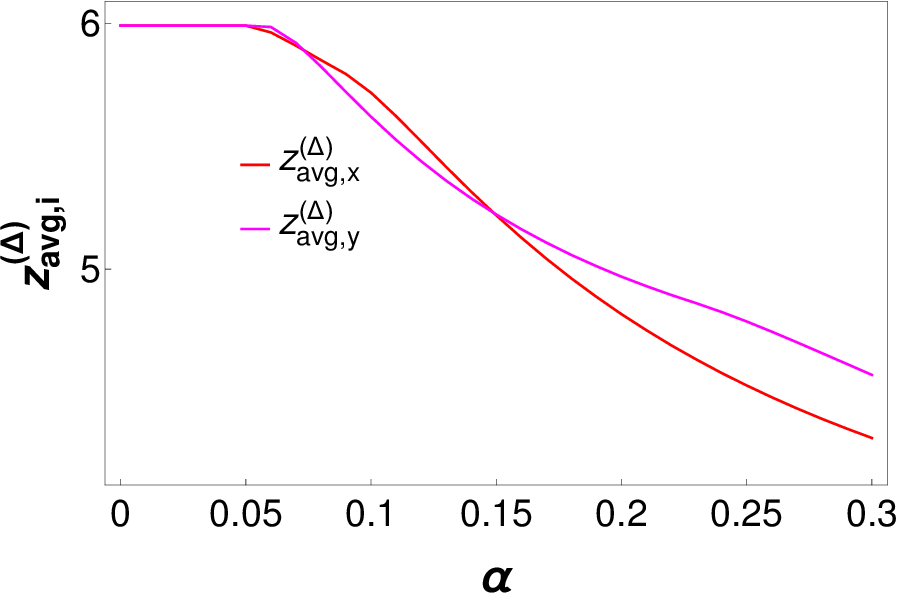}}
     }

\caption{Variation of (a) $p_\mathrm{c}^{(\triangle)}$ and (b) $p_\mathrm{b}^{(\triangle)}$ for a triangular lattice of size $L=2048$ when linear distortion is applied along the $x$ and $y$ directions. The connection threshold is fixed at $d=1.1$. The obtained data points (not shown) are close enough so that the curves appear smooth. In panel (a), the four curves correspond to the site-percolation thresholds $p_\mathrm{c,xx}^{(\triangle)}(\alpha)$ (solid red), $p_\mathrm{c,xy}^{(\triangle)}(\alpha)$ (broken black), $p_\mathrm{c,yx}^{(\triangle)}(\alpha)$ (solid magenta), and $p_\mathrm{c,yy}^{(\triangle)}(\alpha)$ (broken blue). The corresponding curves in panel (b) represent the bond-percolation thresholds $p_\mathrm{b,xx}^{(\triangle)}(\alpha)$, $p_\mathrm{b,xy}^{(\triangle)}(\alpha)$, $p_\mathrm{b,yx}^{(\triangle)}(\alpha)$, and $p_\mathrm{b,yy}^{(\triangle)}(\alpha)$, shown with the same line styles and colors, as defined in the text. The corresponding average coordination numbers $z_\mathrm{avg,x}^{(\triangle)}(\alpha)$ and $z_\mathrm{avg,y}^{(\triangle)}(\alpha)$ are shown in panel (c). A crossing is observed at $\alpha\approx 0.15$ in all three panels.}
\label{fig:tri_comp}
\end{figure*}

\section{Connection criterion and cluster formation mechanism}\label{sec:cluster}
After obtaining a distorted lattice either through scheme (i) or (ii) as described in Sec.~\ref{sec:model}, the occupation process is initiated. Prior to this, however, one needs to specify a connection criterion between the nearest neighbors since nearest-neighbor distances are modified due to distortion. Two adjacent sites are eligible to be connected by a bond if their separation is less than or equal to a prescribed connection threshold $d$.  In case of site percolation, all bonds satisfying the connection criterion ($\delta\le d$) are necessarily occupied, whereas the sites themselves are occupied with probability $p$. In contrast, for bond percolation, all the sites are occupied and the bonds are occupied with probability $p$ only if the connection criterion  is satisfied. The process of occupying a sites (bonds) and the determination of the site (bond) percolation threshold $p_\mathrm{c} (p_\mathrm{b})$ of a finite lattice consists of the following steps:
\begin{enumerate}
\item Begin with a regular lattice of linear size $L$. Identify all the sites (bonds between all the pairs of adjacent sites) with a specific numbering scheme.
\item Generate a linearly distorted lattice by dislocating the sites for a certain value of the distortion parameter $\alpha$ as discussed in Sec.~\ref{sec:model}.
\item Set a value for the connection threshold $d$.
\item Randomly select an empty site (empty bond).
\item Occupy the site (occupy the bond only if the connection criterion is satisfied).
\item Occupied nearest neighboring sites form a cluster if they satisfy connection criterion (occupied adjacent bonds form clusters).
\item After each occupation, put the occupied site (bond) into proper site-cluster (bond-cluster) and check whether a spanning cluster exists.
\item If a spanning cluster does not exist, repeat steps 4-7.
\item Stop when a spanning cluster is found and calculate the fraction $f_\mathrm{c}$ ($f_\mathrm{b}$) of occupied sites (bonds) .
\item Generate another realization with same $\alpha$ and $d$ and repeat steps 4-9.
\item An average of $1000$ such values of $f_\mathrm{c}$ ($f_\mathrm{b}$) gives $p_\mathrm{c}(\alpha,d)$ ($p_\mathrm{b}(\alpha,d)$).
\end{enumerate}
The cluster numbering and spanning analysis have been done by the Newman-Ziff algorithm.

\section{Variation of site and bond percolation thresholds with distortion}\label{sec:main}
This section presents the central results of this work, namely the variation of site and bond percolation thresholds in linearly distorted triangular and square lattices. We denote the site and bond percolation thresholds of a triangular lattice distorted along the $x$ direction by $p^{(\triangle)}_\mathrm{c,x}$ and $p^{(\triangle)}_\mathrm{b,x}$, respectively. The corresponding thresholds for a square lattice distorted along the $x$ direction are denoted by $p^{(\square)}_\mathrm{c,x}$ and $p^{(\square)}_\mathrm{b,x}$. For distortion applied along the $y$ direction, the notation is modified accordingly by replacing the subscript $x$ with $y$.
\begin{figure*}[!htbp]
	  \subfigure[]{\includegraphics[scale=0.40]{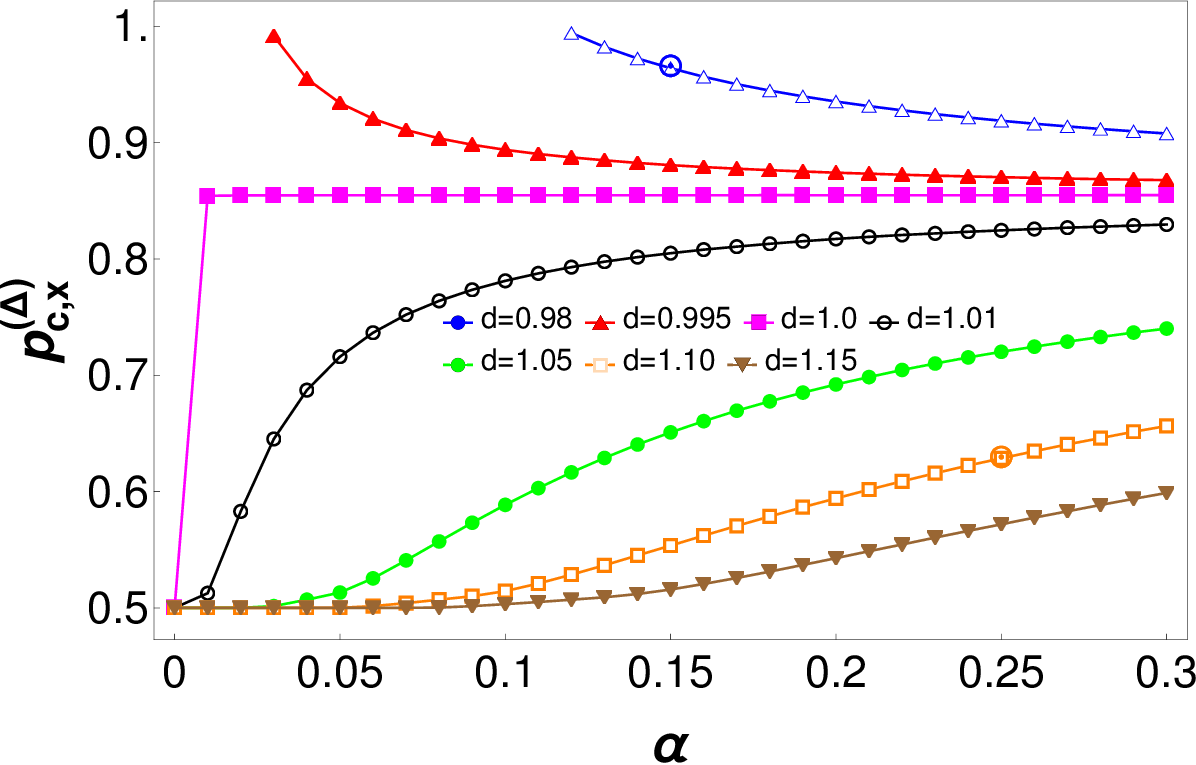}}\hfill
      \subfigure[]{\includegraphics[scale=0.41]{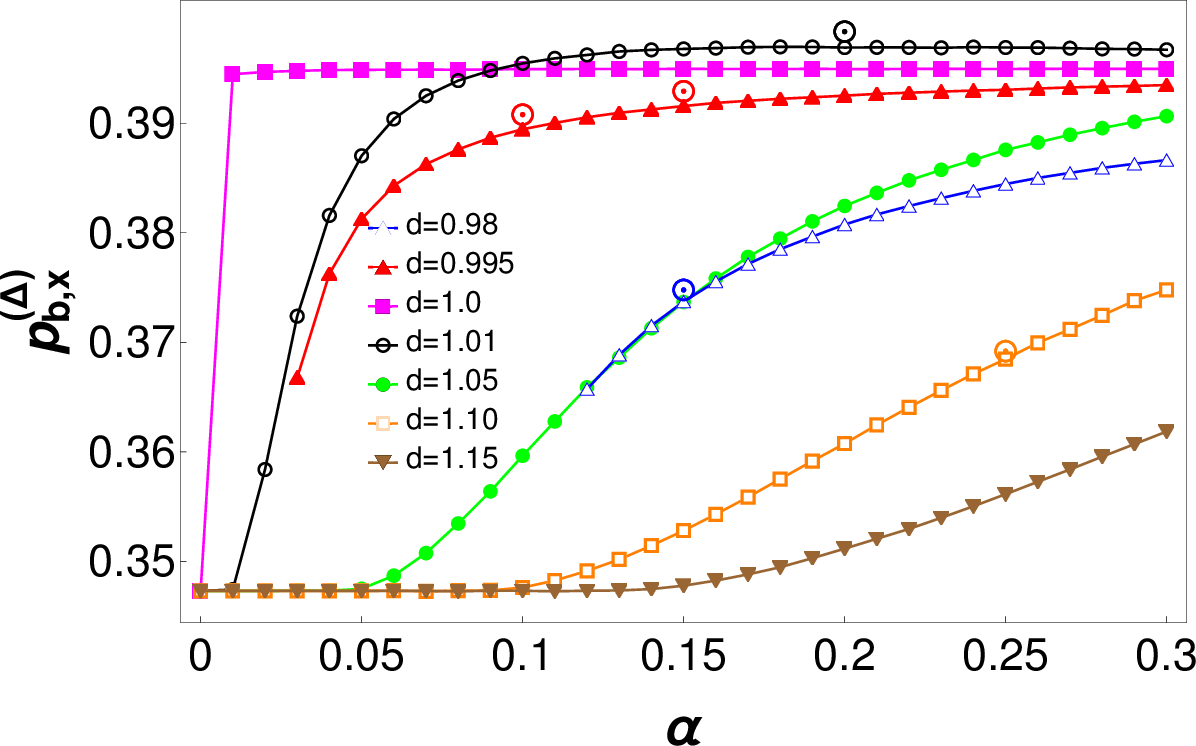}}

\caption{Variation of (a) $p_\mathrm{c,x}^{(\triangle)}$ and (b) $p_\mathrm{b,x}^{(\triangle)}$ for a triangular lattice of size $L=2048$. Each data point represents an average over $1000$ independent realizations. The associated error bars are of the order of $10^{-5}$ and are therefore hidden by the plot markers. The data points are joined by lines as a guide to the eye. Each curve corresponds to a distinct value of $d$ and is indicated by a different color and plot marker. The corresponding percolation thresholds in the thermodynamic limit obtained using Binder cumulant crossings (see Sec. \ref{sec:inf}) for some selected values of $d$  and $\alpha$ are shown by the symbol $\odot$ in the same colors.}
\label{fig:tri_x}
\end{figure*}

\begin{figure}[!htbp]
\includegraphics[scale=0.5]{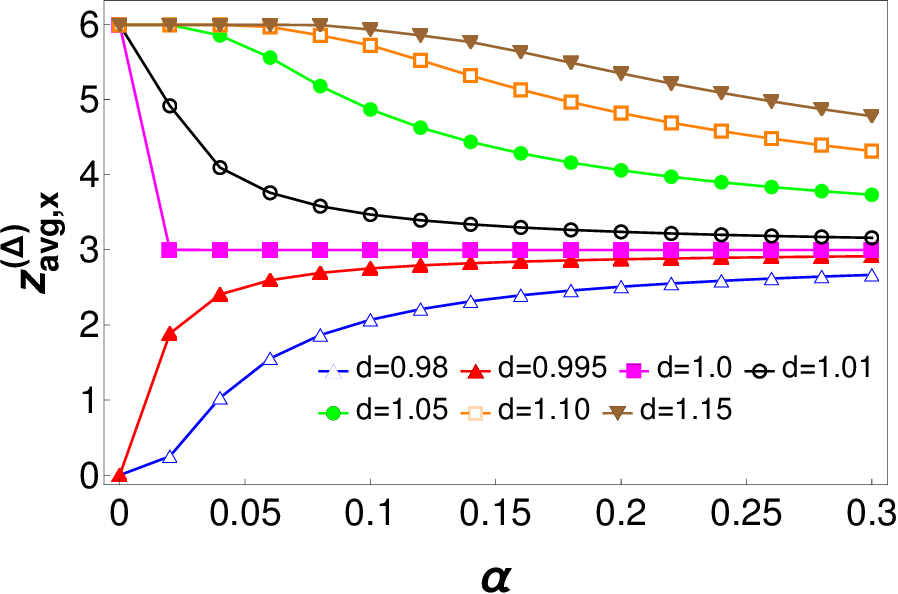}
\caption{Plots of the average coordination number $z_\mathrm{avg,x}^{(\triangle)}$ for the same set of values of $d$ as in Fig.\ \ref{fig:tri_x}. Curves corresponding to identical values of $d$ in the two figures are indicated using the same color and plot symbol. Each data point represents an average over $100$ independent realizations. The associated error bars are of the order of $10^{-5}$ and are therefore hidden by the plot markers. The data points are joined by lines as a guide to the eye.}
\label{fig:tri_x_zavg}
\end{figure}
Why, then, are $p_\mathrm{c,x}^{(\triangle)}(\alpha)$ and $p_\mathrm{c,y}^{(\triangle)}(\alpha)$  [or, $p_\mathrm{b,y}^{(\triangle)}(\alpha)$ and $p_\mathrm{b,y}^{(\triangle)}(\alpha)$] different? We believe that a satisfactory explanation can be provided from Eq.\ \ref{eq:maxmin} and Fig.\ \ref{fig:dist_mech_tri}. The minimum and the maximum nearest neighbor separations are direction dependent, namely $\delta_{\mathrm{min,x}}^{\mathrm{(\triangle)}} \neq \delta_{\mathrm{min,y}}^{\mathrm{(\triangle)}}$ and $\delta_{\mathrm{max,x}}^{\mathrm{(\triangle)}} \neq \delta_{\mathrm{max,y}}^{\mathrm{(\triangle)}}$. As a result, connectivity is established differently when distortion is introduced along the $x$ and $y$ directions leading to distinct percolation thresholds for the triangular lattice.

\subsection{Triangular lattice}\label{sec:tri}
 
Before presenting the detailed results for distorted triangular lattices, an important and striking feature must be discussed. From Eq. \ref{eq:maxmin}, we see that the minimum and maximum nearest-neighbor distances of a distorted triangular lattice differ for $x$- and $y$-distortions, unlike in the square lattice. This difference directly affects connectivity and cluster formation, and can therefore lead to different percolation thresholds for triangular lattices distorted along the $x$ and $y$ directions. In addition, since our analysis is performed on finite lattices, it is necessary to examine spanning along both the $x$ and $y$ directions.

Fig.\ \ref{fig:tri_comp} shows the variation of the (a) site percolation threshold $p_\mathrm{c,ij}^{(\triangle)}(\alpha)$ and (b) bond percolation threshold $p_\mathrm{b,ij}^{(\triangle)}(\alpha)$ for a fixed $d=1.1$. Here the first index $i$ represents the direction of distortion and the second index $j$ represents the direction along which the spanning is checked. The four curves of Fig.\ \ref{fig:tri_comp}(a) represent $p_\mathrm{c,xx}^{(\triangle)}(\alpha)$ (solid black), $p_\mathrm{c,xy}^{(\triangle)}(\alpha)$ (broken red), $p_\mathrm{c,yx}^{(\triangle)}(\alpha)$ (solid blue), and $p_\mathrm{c,yy}^{(\triangle)}(\alpha)$ (broken magenta). 

Fig.\ \ref{fig:tri_comp}(a) clearly shows that $p_\mathrm{c,xx}^{(\triangle)}(\alpha)=p_\mathrm{c,xy}^{(\triangle)}(\alpha)$ and $p_\mathrm{c,yx}^{(\triangle)}(\alpha)=p_\mathrm{c,yy}^{(\triangle)}(\alpha)$. This confirms that $p_\mathrm{c}^{(\triangle)}(\alpha)$ does not depend on whether the spanning is checked along $x$ or $y$ direction, as expected. However, we also observe that $p_\mathrm{c,xx}^{(\triangle)}(\alpha)\neq p_\mathrm{c,yx}^{(\triangle)}(\alpha)$, demonstrating that the percolation threshold \emph{does} depend on the direction of linear distortion. Therefore, the second index is redundant and may be omitted. Henceforth, we denote the site percolation thresholds for distortion along the $x$ and $y$ directions by $p_\mathrm{c,x}^{(\triangle)}(\alpha)$ and $p_\mathrm{c,y}^{(\triangle)}(\alpha)$, respectively. Based on this observation, we conclude that $p_\mathrm{c,x}^{(\triangle)}(\alpha) \neq p_\mathrm{c,y}^{(\triangle)}(\alpha)$. 

A similar behavior observed for bond percolation  (see Fig.~\ref{fig:tri_comp}(b)) although the magnitudes of the variations in $p_\mathrm{b,x}^{(\triangle)}(\alpha)$ and $p_\mathrm{b,y}^{(\triangle)}(\alpha)$ are different, as expected. The separations between the solid and broken line pairs in Fig.~\ref{fig:tri_comp}(b) reduce with lattice size indicating that this difference arises from finite size effects.

A crossing between the curves $p_\mathrm{c,x}^{(\triangle)}(\alpha)$ and $p_\mathrm{c,y}^{(\triangle)}(\alpha)$ in Fig.~\ref{fig:tri_comp}(a), and a similar crossing between $p_\mathrm{b,x}^{(\triangle)}(\alpha)$ and $p_\mathrm{b,y}^{(\triangle)}(\alpha)$ in Fig.~\ref{fig:tri_comp}(b), is observed at $\alpha \approx 0.15$. This indicates that, at this value of $\alpha$, both the site- and bond-percolation thresholds are identical for $x$- and $y$-distorted triangular lattices.

This behavior can be understood from Fig.~\ref{fig:tri_comp}(c), which shows the average coordination numbers $z_\mathrm{avg,x}^{(\triangle)}(\alpha)$ and $z_\mathrm{avg,y}^{(\triangle)}(\alpha)$ for the two types of distortion. As expected, the increase in the percolation thresholds with distortion reflects a reduction in the average coordination number. The crossing of $z_\mathrm{avg,x}^{(\triangle)}(\alpha)$ and $z_\mathrm{avg,y}^{(\triangle)}(\alpha)$ further confirms that the coordination numbers—and hence the effective connectivity—are equal at $\alpha \approx 0.15$.
\subsubsection{Distortion in $x$ direction}
The variation of the site percolation threshold $p_\mathrm{c,x}^{(\triangle)}(\alpha)$ for a large triangular lattice of size $L=2048$, linearly distorted along the $x$ direction, is shown in Fig.\ \ref{fig:tri_x}(a). The distortion is implemented using the mechanism illustrated in Fig.\ \ref{fig:dist_mech_tri}(a). The observed trends indicate that $p_\mathrm{c,x}^{(\triangle)}(\alpha)$ increases with increasing distortion when the connection threshold satisfies $d \geq 1$. This implies that spanning becomes progressively more difficult as distortion increases when the connection threshold is greater than or equal to the lattice constant. In contrast, as evident from the top two curves corresponding to $d<1$, the opposite behavior is observed when the connection threshold is less than the lattice constant: spanning becomes easier with increasing distortion.

For the special case $d=1.0$, a pronounced discontinuous jump is observed, from $p_\mathrm{c,x}^{(\triangle)}=1/2$ in the undistorted lattice to a substantially larger value for an infinitesimally small distortion $\alpha$. More generally, all seven curves in Fig.\ \ref{fig:tri_x}(a), corresponding to seven different values of $d$, exhibit significant variation in $p_\mathrm{c,x}^{(\triangle)}(\alpha)$ in the low-$\alpha$ regime. In the high-distortion regime, however, the behavior becomes more gradual: the curves for (i) $d<1$ decrease slowly, (ii) $d>1$ increase slowly, while (iii) the curve for $d=1$ becomes essentially independent of the distortion strength.

All these features of $p_\mathrm{c,x}^{(\triangle)}(\alpha)$ can be understood from the plots of average coordination number $z_\mathrm{avg,x}^{(\triangle)}(\alpha)$ shown in Fig.\ \ref{fig:tri_x_zavg}. A comparison of Fig.\ \ref{fig:tri_x}(a) and Fig.\ \ref{fig:tri_x_zavg} reveals that the curves corresponding to the same values of $d$ exhibit opposite trends. This is expected since an increase in $z_\mathrm{avg,x}^{(\triangle)}$ facilitates connectivity and reduces the site percolation threshold and vice versa. The discontinuous jump for $d=1.0$ is also observed for $z_\mathrm{avg,x}^{(\triangle)}$ in Fig.\ \ref{fig:tri_x_zavg}.

Fig.\ \ref{fig:tri_x}(b) shows the variation of the bond percolation threshold $p_\mathrm{b,x}^{(\triangle)}(\alpha)$ for the same set of values of $d$. The lattice size and the distortion mechanism are the same as Fig.\ \ref{fig:tri_x}(a). Although the site and bond thresholds show some similarities, we note the following crucial differences
\begin{enumerate}
\item While the site percolation threshold $p_\mathrm{c,x}^{(\triangle)}(\alpha)$ shows a decreasing trend for the two curves with $d<1$ (the top two curves), the bond percolation threshold  $p_\mathrm{b,x}^{(\triangle)}(\alpha)$ always increases with distortion. 
\item By comparing the curves corresponding to $d<1$ in Figs. \ref{fig:tri_x}(b) and \ref{fig:tri_x_zavg}, one observes that $z_\mathrm{avg,x}^{(\triangle)}(\alpha)$ and $p_\mathrm{b,x}^{(\triangle)}(\alpha)$ increase simultaneously. 
\item The curves of $z_\mathrm{avg,x}^{(\triangle)}(\alpha)$ in Fig.\ \ref{fig:tri_x_zavg} are progressively shifted upward as $d$ increases. The corresponding curves of $p_\mathrm{c,x}^{(\triangle)}(\alpha)$ in Fig.\ \ref{fig:tri_x}(a) are placed in the reversed order, reflecting the inverse relationship between coordination number and site threshold. This correspondence is broken for $p_\mathrm{b,x}^{(\triangle)}(\alpha)$. The two curves for $d=0.98$ and $0.995$ do not occupy the topmost positions in Fig.\ \ref{fig:tri_x}(b).
\item There is a crossing between the curves of $p_\mathrm{b,x}^{(\triangle)}(\alpha)$ for $d=1.0$ and $d=1.01$, while no corresponding crossing is observed in $z_\mathrm{avg,x}^{(\triangle)}(\alpha)$ curves. 
\end{enumerate}

All the above  counterintuitive facts establish that, unlike $p_\mathrm{c,x}^{(\triangle)}(\alpha)$, the variation of $p_\mathrm{b,x}^{(\triangle)}(\alpha)$ cannot be explained solely by the variations of $z_\mathrm{avg,x}^{(\triangle)}(\alpha)$. To verify this behavior of $p_\mathrm{b,x}^{(\triangle)}(\alpha)$ of a finite triangular lattice, we have determined the $p_\mathrm{b,x}^{(\triangle)}$ in the thermodynamic limit through the intersections of the Binder cumulant (see Sec.~\ref{sec:inf}) for some selected values of $\alpha$ and $d$. As can be seen from Fig.\ \ref{fig:tri_x}(b), the values corresponding to the infinite and the finite ($L=2048$) lattices are quite close. This confirms the correctness of some of the counterintuitive patterns of $p_\mathrm{b,x}^{(\triangle)}(\alpha)$ and suggests that the cluster formation and the spanning mechanisms at a microscopic level are much more complex and non-trivial for bond percolation in a triangular lattice distorted in the $x$ direction.

The curves corresponding to $d=1.0$ in Figs.~\ref{fig:tri_x}(a), \ref{fig:tri_x}(b), and \ref{fig:tri_x_zavg} exhibit a distinctive behavior and therefore merit special attention. As seen from Fig.~\ref{fig:tri_x_zavg}, the average coordination number $z_\mathrm{avg,x}^{(\triangle)}(\alpha)$ drops abruptly from $6$ at $\alpha=0$ to $3$ for an infinitesimally small distortion. A closer inspection of Fig.~\ref{fig:dist_mech_tri}(a) shows that random displacements of sites along the $\pm x$ directions cause both horizontal and diagonal bonds to be stretched or compressed with equal probability. Therefore, when the connection threshold equals the lattice constant, approximately half of the nearest-neighbor bonds fail to satisfy the connection criterion. As a result, the average coordination number is halved. This situation persists for all nonzero values of $\alpha$; consequently, $z_\mathrm{avg,x}^{(\triangle)}(\alpha)$ remains constant at $3$.

It is therefore not surprising that the site percolation threshold $p_\mathrm{c,x}^{(\triangle)}(\alpha)$ remains constant for all $\alpha>0$. Its value ($\sim 0.854$), however, is slightly higher than those of other lattices with coordination number $3$ \cite{SudingZiff1999, Jacobsen_2014}. This is expected, since the partial loss of connections requires a larger fraction of oocupied sites to achieve spanning. In contrast, the bond percolation threshold $p_\mathrm{b,x}^{(\triangle)}(\alpha)$ also remains constant at a value $\sim 0.3945$, which is significantly lower than those of other lattices with the same coordination number \cite{Jacobsen_2014,Scullard2020} . This difference arises from the reduced bond-occupation randomness in the present model:  a bond can only be occupied if and only if it satisfies the connection criterion. A similar distinction between site-bond percolation and percolation with distortion has been noted earlier \cite{Sayantan2}.

\begin{figure*}[!htbp]
	  \subfigure[]{\includegraphics[scale=0.53]{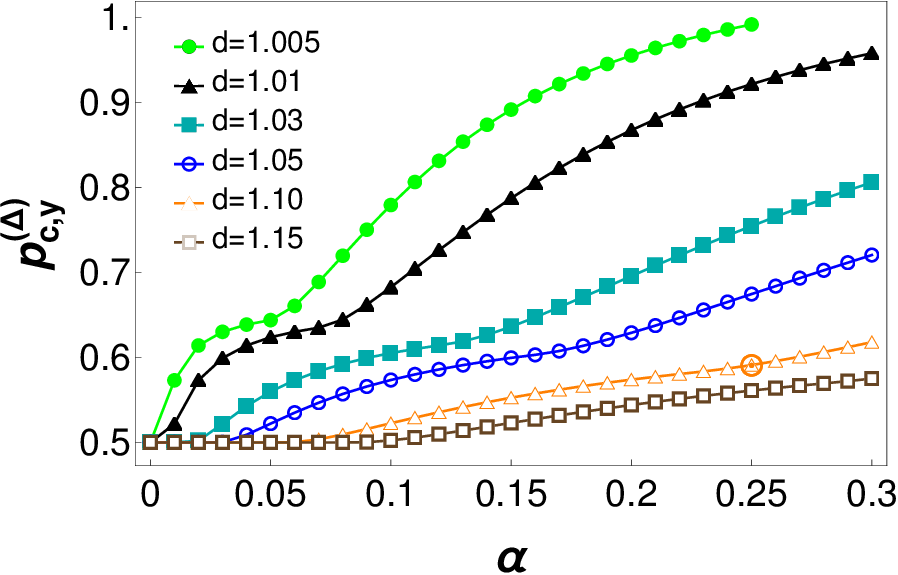}}\hfill
      \subfigure[]{\includegraphics[scale=0.55]{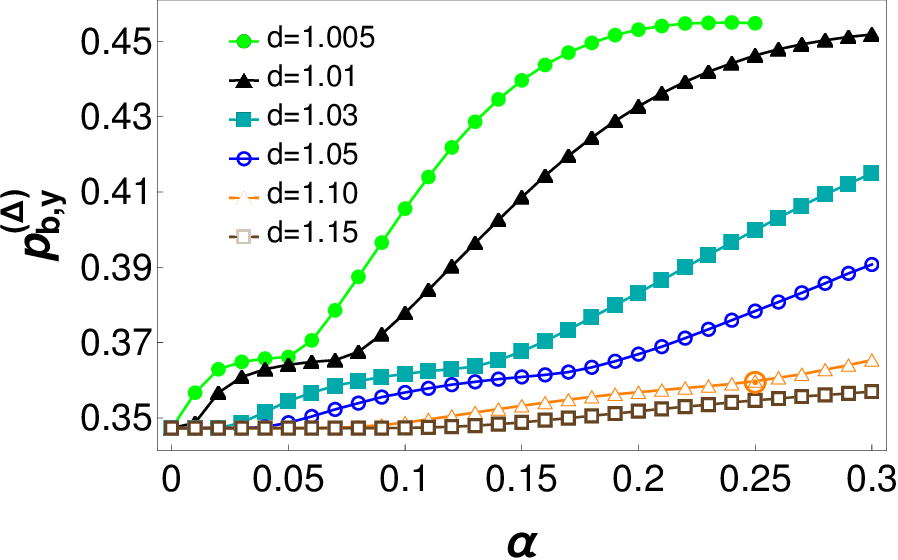}}

\caption{Variation of (a) $p_\mathrm{c,y}^{(\triangle)}$ and (b) $p_\mathrm{b,y}^{(\triangle)}$ for a triangular lattice of size $L=2048$. Each data point represents an average over $1000$ independent realizations. The associated error bars are of the order of $10^{-5}$ and are therefore hidden by the plot markers. The data points are joined by lines as a guide to the eye. Each curve corresponds to a distinct value of $d$ and is indicated by a different color and plot marker. The corresponding percolation thresholds in the thermodynamic limit obtained using Binder cumulant crossings (see Sec. \ref{sec:inf}) or $d=1.10$ and $\alpha=0.25$ are shown by the symbol $\odot$ in the same color.}
\label{fig:tri_y}
\end{figure*}

\begin{figure}[!htbp]
\includegraphics[scale=0.5]{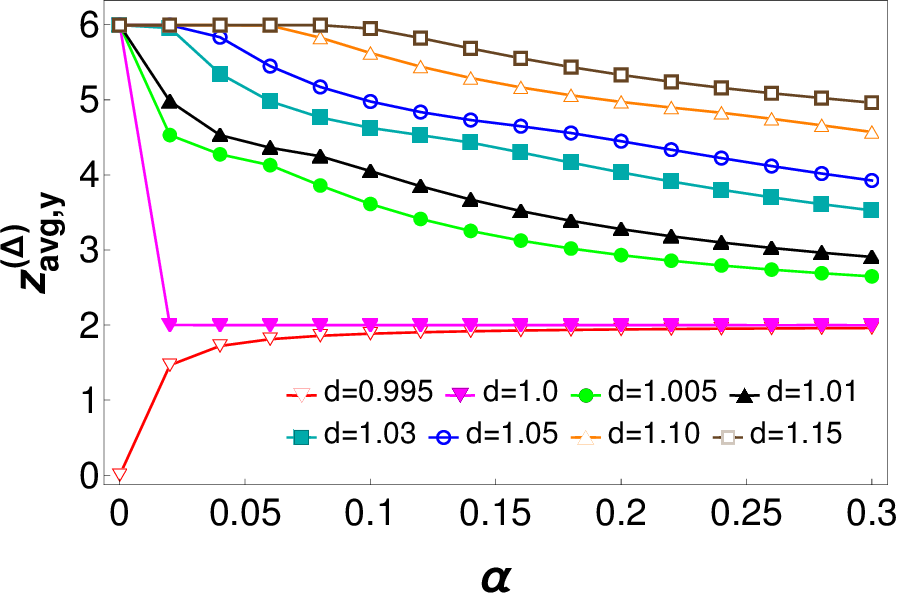}
\caption{Plots of $z_\mathrm{avg,y}^{(\triangle)}$ for the same set of values of $d$ as in Fig.\ \ref{fig:tri_y}. Curves corresponding to identical values of $d$ in the two figures are indicated using the same color and plot symbol. Each data point represents an average over $100$ independent realizations. The associated error bars are of the order of $10^{-5}$ and are therefore hidden by the plot markers. The data points are joined by lines as a guide to the eye.}
\label{fig:tri_y_zavg}
\end{figure}

\subsubsection{Distortion in $y$ direction}
As discussed earlier, the patterns of variations of both site and bond percolation thresholds with the distortion parameter are different when linear distortion is employed in the $x$ and $y$ directions.  The variations of the site percolation threshold [$p_\mathrm{c,y}^{(\triangle)}(\alpha)$] and the bond percolation threshold [$p_\mathrm{b,y}^{(\triangle)}(\alpha)$] of a triangular lattice of size $L=2048$ linearly distorted in the $y$ direction are shown in Fig.\ \ref{fig:tri_y}(a) and \ref{fig:tri_y}(b) respectively. Fig.\ \ref{fig:tri_y_zavg} shows the corresponding plots of the average coordination number $z_\mathrm{avg,y}^{(\triangle)}(\alpha)$ for the same set of values of $d$. The following features emerge from these plots.

\begin{enumerate}
\item A careful inspection of Fig.~\ref{fig:dist_mech_tri}(b) reveals that, when the sites of a triangular lattice are dislocated in the $y$ direction, the horizontal bonds are always stretched and the diagonal bonds are equally likely to be stretched or compressed. As a result, when the connection threshold is set equal to the lattice constant, the average coordination number drops from $6$ to $2$ when distortion is turned on. The plot of $z_\mathrm{avg,y}^{(\triangle)}(\alpha)$ corresponding to $d=1.0$ in Fig.~\ref{fig:tri_y_zavg} explicitly demonstrates the above fact. Since this value is too low to generate a sufficient number of connections, no spanning cluster is found for $d=1$. As observed previously for other distortion schemes \cite{Bishnu1}, the value of $z_\mathrm{avg}$ has to be significantly larger than $2$ to give rise to a spanning cluster. The curve of $z_\mathrm{avg,y}^{(\triangle)}(\alpha)$ for $d=0.995$ stays even below that. Consequently, the triangular lattice linearly distorted linearly in the $y$ direction cannot percolate when the connection threshold is less than or equal to the lattice constant. 

\item Although the magnitude and range of variation of the site percolation threshold are much higher, the variation patterns of $p_\mathrm{c,y}^{(\triangle)}(\alpha)$ and $p_\mathrm{b,y}^{(\triangle)}(\alpha)$ are very similar. All the curves start from the regular $p_\mathrm{c}$ or $p_\mathrm{b}$ value for the triangular lattice, as they should. When the distortion is large enough so that $\delta_{\mathrm{max,y}}^{\mathrm{(\triangle)}}>d$, some bonds fail to satisfy the connection criterion and the percolation threshold starts to increase. In general, spanning becomes difficult as distortion increases.

\item After the initial rise, a plateau-like region is observed, where the percolation threshold essentially stays constant with $\alpha$. A steady increase is seen thereafter. This plateau is more prominent for low $d$-values, for which the bond percolation threshold becomes almost constant in the high-distortion regime. 

\end{enumerate}

\begin{figure}[!htbp]
\includegraphics[scale=0.53]{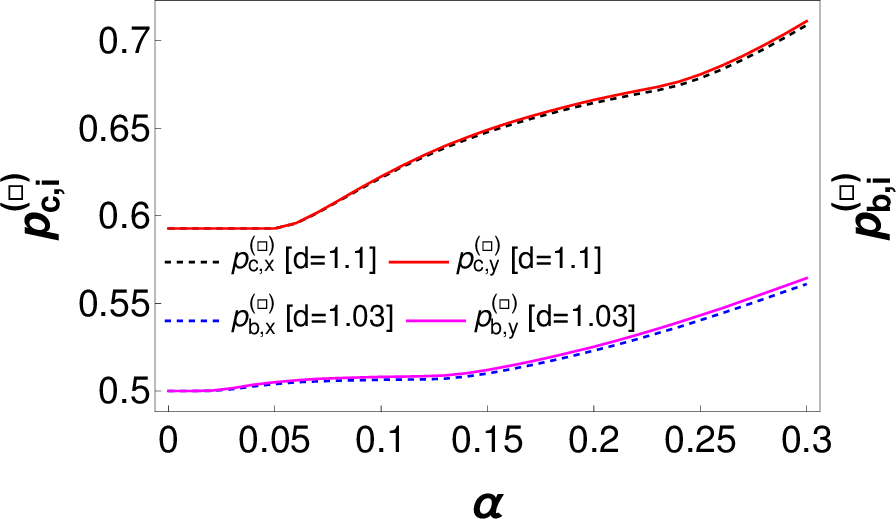}
\caption{Plots of the site-percolation thresholds (top) $p_\mathrm{c,x}^{(\square)}(\alpha)$ and $p_\mathrm{c,y}^{(\square)}(\alpha)$ for $d=1.1$, and the bond-percolation thresholds (bottom) $p_\mathrm{b,x}^{(\square)}(\alpha)$ and $p_\mathrm{b,y}^{(\square)}(\alpha)$ for $d=1.03$, demonstrating that the thresholds are insensitive to the direction of linear distortion.}
\label{fig:squ_comp}
\end{figure}

\begin{figure*}[!htbp]
	  \subfigure[]{\includegraphics[scale=0.54]{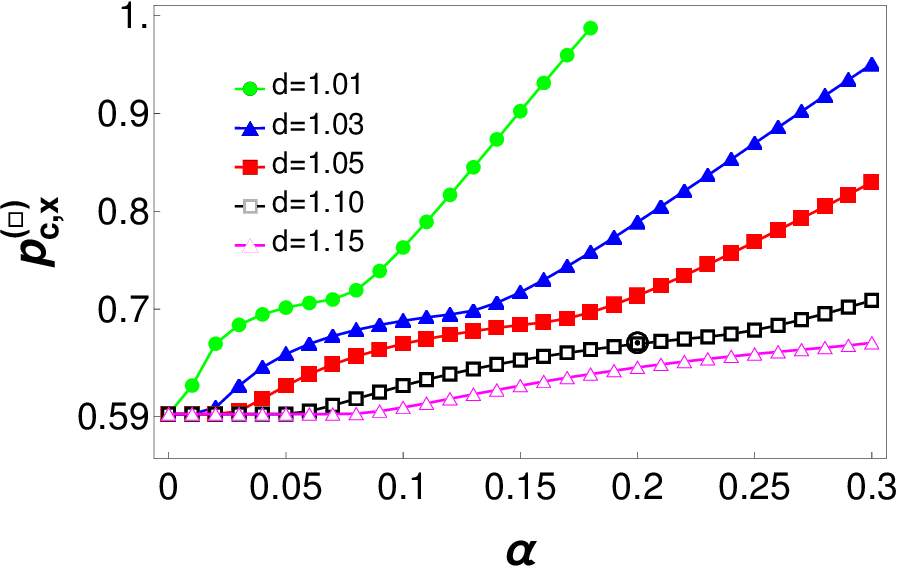}}\hfill
      \subfigure[]{\includegraphics[scale=0.54]{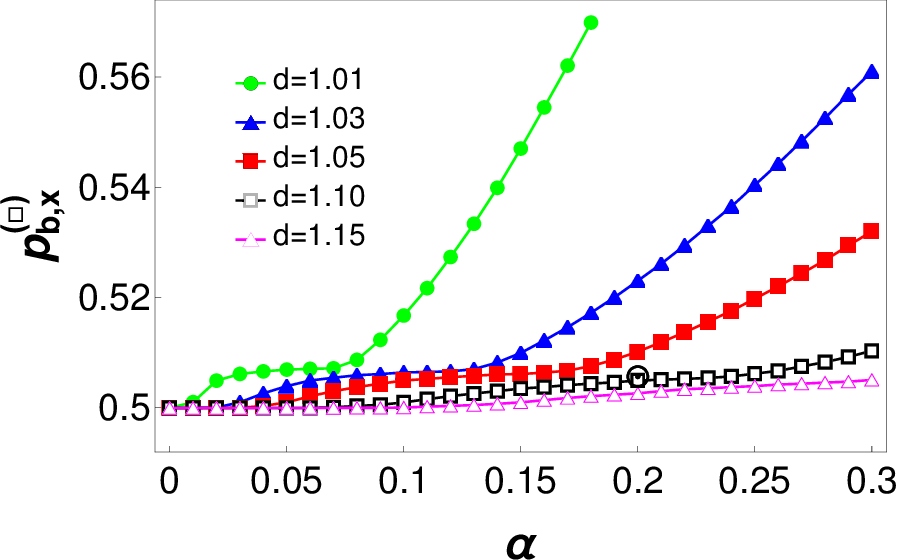}}

\caption{Variation of (a) $p_\mathrm{c,x}^{(\square)}$ and (b) $p_\mathrm{b,x}^{(\square)}$ for a square lattice of size $L=2048$. Each data point represents an average over $1000$ independent realizations. The associated error bars are of the order of $10^{-5}$ and are therefore hidden by the plot markers. The data points are joined by lines as a guide to the eye. Each curve corresponds to a distinct value of $d$ and is indicated by a different color and plot marker. The corresponding percolation thresholds in the thermodynamic limit obtained using Binder cumulant crossings (see Sec. \ref{sec:inf}) for $d=1.10$ and $\alpha=0.2$ are shown by the symbol $\odot$ in the same color.}
\label{fig:squ_x}
\end{figure*}

\begin{figure}[!htbp]
\includegraphics[scale=0.53]{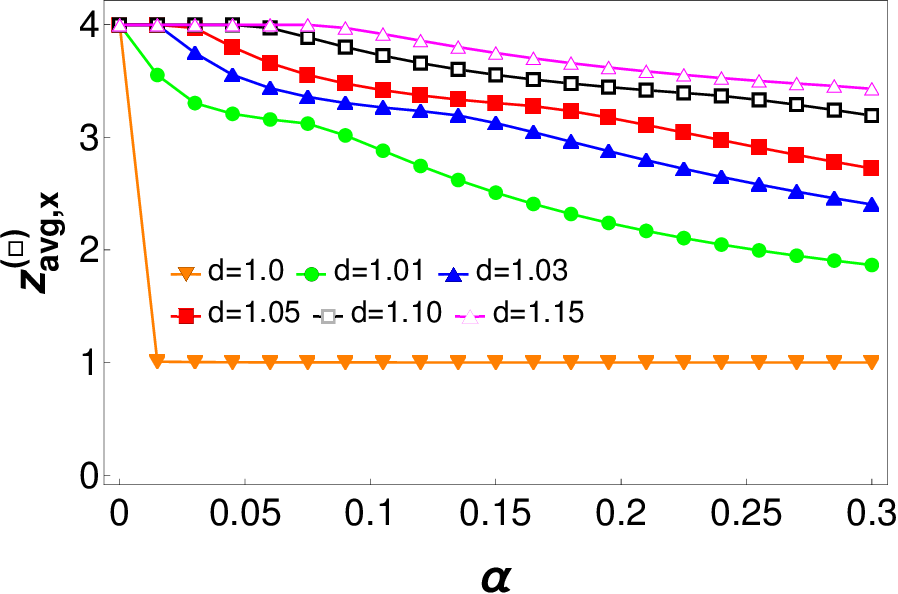}
\caption{Plots of $z_\mathrm{avg,x}^{(\square)}$ for the same set of values of $d$ as in Fig.\ \ref{fig:squ_x}. Curves corresponding to identical values of $d$ in the two figures are indicated using the same color and plot symbol. Each data point represents an average over $100$ independent realizations. The associated error bars are of the order of $10^{-5}$ and are therefore hidden by the plot markers. The data points are joined by lines as a guide to the eye.}
\label{fig:squ_x_zavg}
\end{figure}

\subsection{Square lattice}\label{sec:squ}
For finite-sized square lattices, the two site percolation thresholds $p_\mathrm{c,x}^{(\square)}(\alpha)$ and $p_\mathrm{c,y}^{(\square)}(\alpha)$, and the two bond percolation thresholds $p_\mathrm{b,x}^{(\square)}(\alpha)$ and $p_\mathrm{b,y}^{(\square)}(\alpha)$ are very close as shown in Fig.\ \ref{fig:squ_comp}. The slight difference diminishes as the lattice size increases and therefore they coincide in the thermodynamic limit. This is expected since $\delta_{\mathrm{min,x}}^{\mathrm{(\square)}}=\delta_{\mathrm{min,y}}^{\mathrm{(\square)}}=1-2\alpha$ and $\delta_{\mathrm{max,x}}^{\mathrm{(\square)}}= \delta_{\mathrm{max,y}}^{\mathrm{(\square)}}=1+2\alpha$. In a square lattice, therefore, linear distortion along different directions does not change the connectivity pattern between the sites leading to the identical percolation thresholds: $p_\mathrm{c,x}^{(\square)}(\alpha)=p_\mathrm{c,y}^{(\square)}(\alpha)$ and $p_\mathrm{b,x}^{(\square)}(\alpha)=p_\mathrm{b,y}^{(\square)}(\alpha)$.  

It is thus sufficient, for a square lattice, to present the site and bond percolation thresholds for linear distortion in any one direction. Fig.\ \ref{fig:squ_x}(a), \ref{fig:squ_x}(b), and \ref{fig:squ_x_zavg} show the plots of $p_\mathrm{c,x}^{(\square)}(\alpha)$, $p_\mathrm{b,x}^{(\square)}(\alpha)$, and $z_\mathrm{avg,x}^\mathrm{(\square)}$ respectively, for the same set of values of the connection threshold $d$. The salient features of these plots are the following.
\begin{enumerate}
\item A linearly distorted square lattice cannot span when $d\le 1$. It can be understood from Fig.~\ref{fig:dist_mech_squ}(a) and (b) that the bonds parallel to the direction of linear distortion are equally likely to be stretched or compressed, while the bonds perpendicular to distortion direction are always stretched. Therefore, out of four bonds around a site, only one satisfies the connection criterion on an average when $d=1.0$. Fig.~\ref{fig:squ_x_zavg} reveals that $z_\mathrm{avg,x}^\mathrm{(\square)}$ abruptly drops from $4$ to $1$ when $\alpha$ changes from $0$ to any infinitesimal value. As observed earlier, it is not possible for a distorted lattice to develop enough connections for spanning if the average coordination number drops to $2$ or less.
\item Site and bond percolation thresholds both increase with distortion. All curves start from the threshold of the undistorted lattice ($\approx 0.59$ for site and $\approx 0.5$ for bond) increase initially, form a plateau, and then increase steadily thereafter. This pattern is particularly prominent for low values of $d$.
\item As observed for all lattice types, the magnitude of the change in the threshold for site percolation is much larger than that for bond percolation.
\end{enumerate}

\begin{figure}[!htbp]
\includegraphics[scale=0.35]{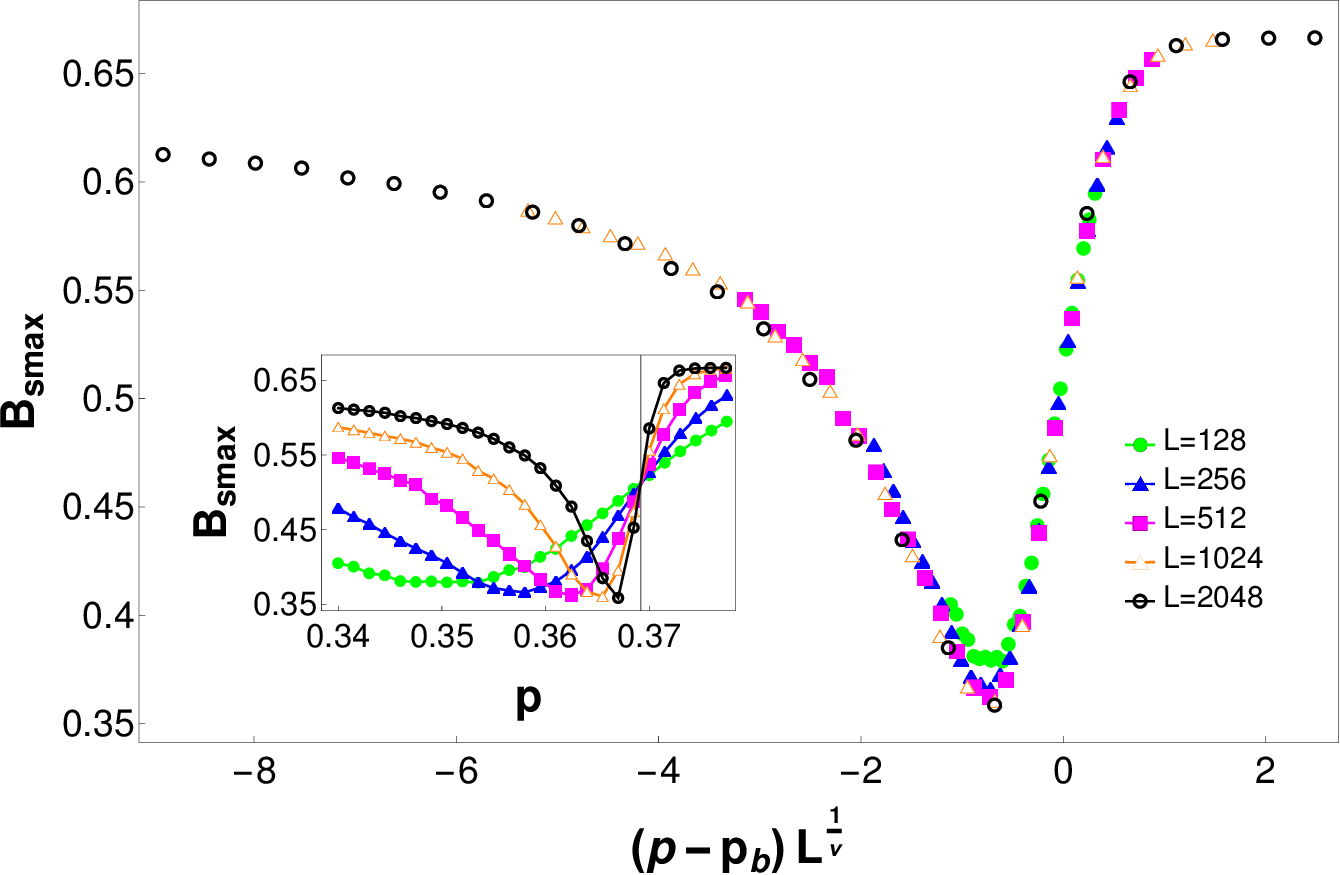}
\caption{Demonstration of finite-size scaling through Binder cumulant plots for $x$-distorted triangular lattices of various system sizes at $d=1.1$ and $\alpha=0.25$.
The inset shows the determination of the bond-percolation threshold in the thermodynamic limit from the intersections of the Binder cumulant curves for five different system sizes. The data points are joined by lines as a guide to the eye. The vertical line indicates the estimated threshold $p_\mathrm{b,x}^{(\triangle)}(d=1.1,\alpha=0.25)=0.36919$ (see Table~\ref{table1}). This value is used to rescale the horizontal axis as $(p-p_\mathrm{b})L^{1/\nu}$, resulting in an excellent data collapse, as shown in the main panel. Each data point represents an average over $10^5$ independent realizations. The associated error bars are of the order of $5 \times 10^{-3}$ and are smaller than the plot markers.}
\label{fig:tri_binder}
\end{figure}

\begin{table}
\centering
\caption{Site and bond percolation thresholds for distorted infinite lattices obtained from intersections of Binder cumulant for both triangular and square lattices with selected values of $d$ and $\alpha$.}
\label{table1}
\renewcommand{\arraystretch}{1.6} 
\begin{ruledtabular}
\begin{tabular}{cccc}
Percolation threshold & $d$ & $\alpha$ & Numerical estimate \\ 
\hline
$p_\mathrm{c,x}^{(\triangle)}$ & 0.98  & 0.15 & 0.96604(2) \\ 
$p_\mathrm{c,x}^{(\triangle)}$ & 1.1   & 0.25 & 0.62979(6) \\
$p_\mathrm{c,y}^{(\triangle)}$ & 1.1   & 0.25 & 0.59071(5) \\
$p_\mathrm{b,x}^{(\triangle)}$ & 0.98  & 0.15 & 0.37478(4) \\
$p_\mathrm{b,x}^{(\triangle)}$ & 0.995 & 0.10 & 0.39079(5) \\
$p_\mathrm{b,x}^{(\triangle)}$ & 0.995 & 0.15 & 0.39285(3) \\
$p_\mathrm{b,x}^{(\triangle)}$ & 1.01  & 0.20 & 0.39828(5) \\
$p_\mathrm{b,x}^{(\triangle)}$ & 1.1   & 0.25 & 0.36919(4) \\
$p_\mathrm{b,y}^{(\triangle)}$ & 1.1   & 0.25 & 0.35946(3) \\
$p_\mathrm{c,x}^{(\square)}$   & 1.1   & 0.20 & 0.66560(6) \\
$p_\mathrm{c,y}^{(\square)}$   & 1.1   & 0.20 & 0.66560(7) \\
$p_\mathrm{b,x}^{(\square)}$   & 1.1   & 0.20 & 0.50565(4) \\
$p_\mathrm{b,y}^{(\square)}$   & 1.1   & 0.20 & 0.50568(5) \\
\end{tabular}
\end{ruledtabular}
\end{table}

\subsection{Thresholds in the thermodynamic limit}\label{sec:inf}
All the values of the percolation thresholds in Secs. \ref{sec:tri} and \ref{sec:squ} correspond to lattices of size $L=2048$ and are averaged over $1000$ independent configurations. In this section, we determine the corresponding thresholds in the thermodynamic limit for a selected set of values of $\alpha$ and $d$. We use the standard method of Binder cumulant intersections. The Binder cumulant associated with the largest cluster is defined as
\begin{equation}
B_\mathrm{smax}(p,L)=1-\frac{\langle [S_\mathrm{max}(p,L)]^4\rangle}{3\langle [S_\mathrm{max}(p,L)]^2\rangle^2},
\end{equation}
where $S_\mathrm{max}(p,L)$ denotes the size of the largest cluster in a lattice of linear size $L$ at occupation probability $p$. The angular brackets $\langle \cdots \rangle$ represent an average taken over a large number of independent lattice realizations; in the present study, $10^5$ realizations were used. The percolation threshold in the thermodynamic limit is then estimated from the common intersection point (within numerical accuracy) of the Binder cumulant curves $B_\mathrm{smax}(p)$ obtained for different system sizes. 

The inset of Fig.~\ref{fig:tri_binder} shows the determination of $p_\mathrm{b,x}^{(\triangle)}(d=1.1,\alpha=0.25)$ in the thermodynamic limit from the Binder cumulant $B_\mathrm{smax}(p)$ for five system sizes, $L=128, 256, 512, 1024$, and $2048$. The threshold is estimated by averaging the occupation probability $p$ over all ten intersection points. The validity of this estimate is confirmed by the excellent data collapse in the main panel of Fig.~\ref{fig:tri_binder}, obtained by rescaling the horizontal axis as $(p-p_\mathrm{b})L^{1/\nu}$. The critical exponent $\nu$ is taken to be $4/3$, consistent with the assumption that the system belongs to the same universality class as regular percolation, as established in our earlier works \cite{Sayantan1, Sayantan2, Bishnu1}.

Table \ref{table1} lists selected site and bond percolation thresholds of $x$- and $y$-distorted triangular and square lattices obtained through the above procedure. All these values are included in Figs. \ref{fig:tri_x}, \ref{fig:tri_y}, \ref{fig:squ_x}. It is evident that the thresholds for the finite-sized ($L=2048$) lattices are quite close to the corresponding thermodynamic-limit values (marked by a different symbol in the same color).

\section{Critical connection threshold}\label{sec:dc}
\begin{figure*}[!htbp]
	  \subfigure[]{\includegraphics[scale=0.53]{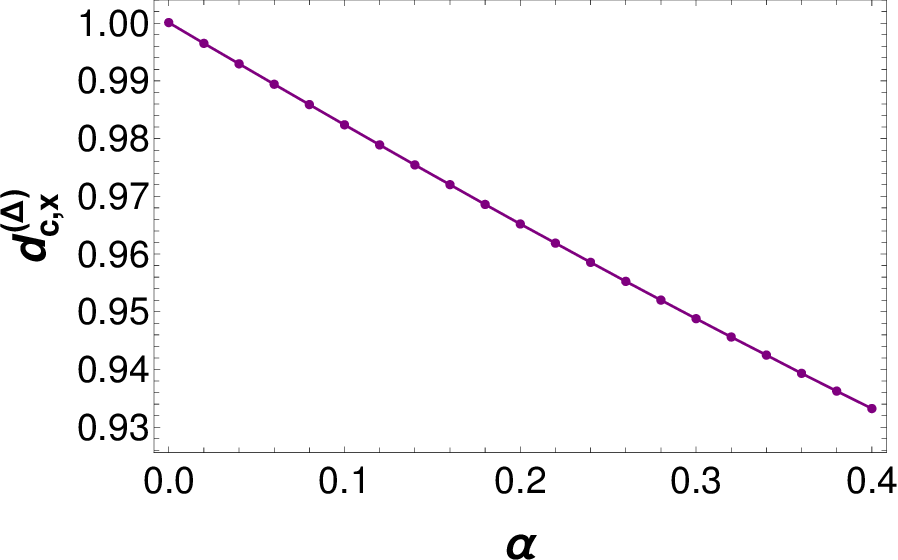}}\hfill
      \subfigure[]{\includegraphics[scale=0.47]{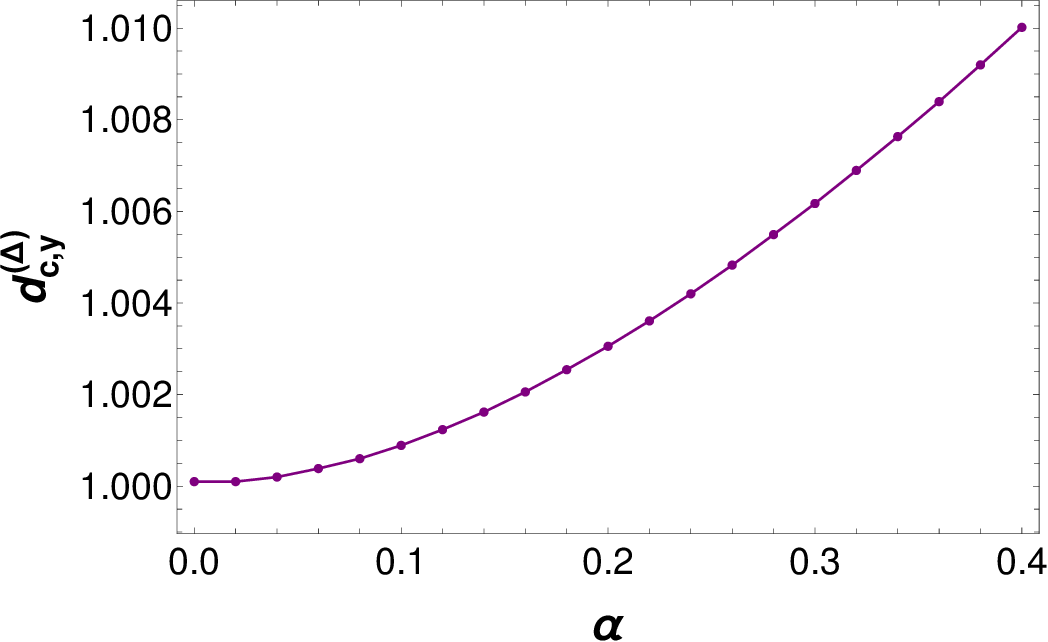}}

\caption{Critical connection thresholds for a triangular lattice of size $L=2048$ subjected to linear distortion: (a) $d_\mathrm{c,x}^{(\triangle)}$ for distortion along the $x$ direction and (b) $d_\mathrm{c,y}^{(\triangle)}$ for distortion along the $y$ direction. Each point is obtained by averaging over $1000$ independent realizations. The error bars are of the order of $10^{-5}$ and are therefore smaller than the plot symbols. Lines are drawn between data points to guide the eye.}
\label{fig:tri_dc}
\end{figure*}

\begin{figure}[!htbp]
\includegraphics[scale=0.53]{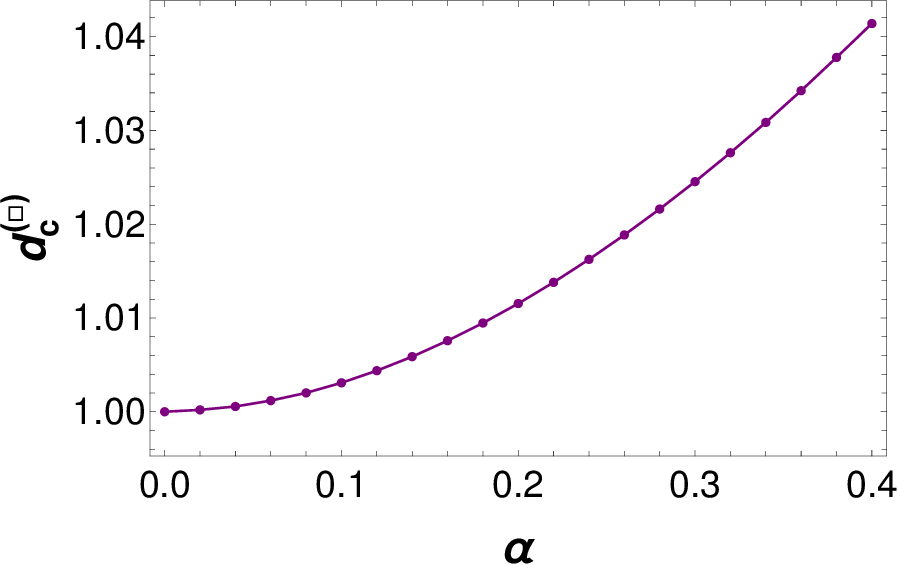}
\caption{Critical connection threshold $d_\mathrm{c}^{(\square)}$ for a linearly distorted square lattice of size $L=2048$. The results are identical for distortion applied in either direction. Each point is obtained by averaging over $1000$ independent realizations. The error bars are of the order of $10^{-5}$ and are therefore smaller than the plot symbols. Lines are drawn between data points to guide the eye.}
\label{fig:squ_dc}
\end{figure}
In our previous article \cite{Bishnu1}, we introduced a quantity called the critical connection threshold $d_\mathrm{c}$ to gain more insight into the cluster building and spanning mechanisms. It is defined as the minimum value of the connection threshold $d$ to ensure spanning when all sites and all allowed bonds are occupied. This quantity explicitly depends on $\alpha$. Therefore, to determine $d_\mathrm{c}$, one first needs to (i) set a value for $\alpha$, (ii) generate a distorted lattice accordingly, (iii) set a low value of $d$ and occupy all the sites and allowed bonds; the value of $d$ should be low enough that no spanning cluster exists, (iv) gradually increase $d$ so that some new bonds are occupied (v) for each value of $d$, check for the existence of a spanning site cluster (for site percolation) or a spanning bond cluster (for bond percolation), (vi) mark the first instance of spanning and record the corresponding value of $d$ (vii) an average over many such values of $d$, obtained by repeating the steps (iii)-(vi), yields an estimate of $d_\mathrm{c}(\alpha)$.

It is worth emphasizing that the critical connection thresholds are exactly the same for site and bond percolation. This important conclusion implies that when all sites and all allowed bonds are occupied, the same minimum value of $d$ triggers the formation of a spanning site-cluster and a spanning bond-cluster.

The critical connection thresholds $d_\mathrm{c,x}^{(\triangle)}(\alpha)$ and $d_\mathrm{c,y}^{(\triangle)}(\alpha)$ of a triangular lattice distorted in the $x$ and $y$ directions are shown in Fig.\ \ref{fig:tri_dc}(a) and (b) respectively. The patterns are markedly different when distortion is applied in $x$ and $y$ directions. While $d_\mathrm{c,x}^{(\triangle)}(\alpha)$ decreases linearly from $1$ as distortion increases, $d_\mathrm{c,y}^{(\triangle)}(\alpha)$ increases parabolically. The impact of distortion is significantly more pronounced for $x$-distortion as the magnitude of the change is larger in this case. This highlights the strong sensitivity of the critical connection threshold to the direction of distortion in case of a linearly distorted triangular lattice.

Owing to the symmetry, there is no directional dependence in the critical connection threshold of a linearly distorted square lattice, so $d_\mathrm{c,x}^{(\square)}(\alpha)=d_\mathrm{c,y}^{(\square)}(\alpha)=d_\mathrm{c}^{(\square)}(\alpha)$. Fig.\ \ref{fig:squ_dc} shows a parabolic increase of $d_\mathrm{c}^{(\square)}(\alpha)$, similar to $d_\mathrm{c,x}^{(\triangle)}(\alpha)$, but with a larger magnitude.
\section{Summary}\label{sec:summary}
We have studied the impact of linear distortion on site and bond percolation in triangular and square lattices using extensive numerical simulations. We have considered two distortion schemes: (i) all the lattice points are distorted along the $\pm x$ direction and (ii) all the lattice points are distorted along the $\pm y$ direction. The points are displaced from their regular positions by a random amount within the interval $[-\alpha, \alpha]$. After dislocation, two originally neighboring points can remain as neighbors only if their separation is less than the connection threshold $d_\mathrm{c}$. We summarize our findings below.
\begin{enumerate}
\item For the triangular lattice, we obtain significantly different results when the distortion is employed along the $x$ and $y$ directions. This reflects the intrinsic anisotropy of the lattice with respect to these two directions, which is revealed by the distortion. The directional dependence of the site- and bond-percolation thresholds, as well as the critical connection threshold, can be understood from the differences in the minimum and maximum bond lengths arising under distortion along different directions.

\item The case of $x$-distortion in triangular lattice appears to be more enigmatic. The behavioral patterns of site and bond percolation thresholds are noticeably different. Variation of the bond percolation threshold is particularly nontrivial and cannot be readily explained from the average coordination number plots.

\item Notably, the bond percolation threshold of a linearly distorted triangular lattice remains unusually low compared to that of regular lattices with similar coordination, indicating that geometric constraints induced by linear distortion fundamentally alter the bond-percolation mechanism.

\item The results of the linearly distorted square lattice follow the expected scenario. Here, the linear distortions in the two directions are equivalent due to symmetry. Consequently, there is no difference in the behavior of any quantity for $x$-distortion and $y$-distortion in a square lattice.

\item When the connection threshold exceeds the lattice constant, all site- and bond-percolation threshold curves of both lattices originate from their undistorted values and begin to increase only beyond a certain value of the distortion parameter $\alpha$. This can be understood as follows. When the connection threshold $d$ is greater than the maximum nearest-neighbor distance, \emph{all} bonds satisfy the connection criterion, and the system effectively does not ``feel'' the disorder. Since the maximum nearest-neighbor distance increases with $\alpha$ (see Eq.~\ref{eq:maxmin}), it eventually exceeds $d$ at sufficiently large distortion, at which point some bonds begin to break, leading to an increase in the percolation threshold.

\item Another general feature of all the percolation threshold curves is that the initially strong variations give way to a monotonic asymptotic behavior. At large distortion, the percolation thresholds exhibit weak dependence on $\alpha$, as the connectivity structure becomes effectively saturated. For the triangular lattice, both site- and bond-percolation thresholds approach nearly constant values at moderate distortion when the connection threshold is close to the lattice constant. The asymptotic thresholds depend on the connection threshold (being higher for smaller connection thresholds), as well as on the lattice type and the nature of percolation (site or bond).

\item The critical connection threshold is identical for site and bond percolation, indicating that when all sites and geometrically allowed bonds are occupied, the onset of spanning is governed solely by geometric connectivity. For a square lattice, it is independent of the direction of linear distortion. In contrast, for a triangular lattice, the critical connection threshold exhibits qualitatively different behavior when distortion is applied along the $x$ and $y$ directions.
\end{enumerate} 

\section{Acknowledgment}
This work is funded by Anusandhan National Research Foundation (ANRF), Department of Science and Technology, Government of India. The project file number is SUR/2022/002345. SM gratefully acknowledges financial support through a National Post Doctoral Fellowship from  ANRF under  project file no. PDF/2023/002952. The authors thank Abdur Rashid Miah for useful discussion. The computation facilities availed at the Department of Physics, University of Gour Banga, Malda are gratefully acknowledged.

\end{document}